\documentstyle[epsfig]{aipproc}

\def\ltap{\raisebox{-.4ex}{\rlap{$\sim$}} \raisebox{.4ex}{$<$}}
\def\rts{\sqrt s}

\def\anti{\overline}

\def\mev{~{\rm MeV}}

\def\anti{\overline}

\def\mt{m_t}

\def\alt{\ltap}

\begin{document}
\noindent \hspace*{3.4in}{\bf IUHET-383}\\
\hspace*{3.4in}{\bf OHSTPY-HEP-E-98-003}\\
\hspace*{3.4in}{\bf February 1998}
\title{Top Quark Physics at Muon\\ and other Future 
Colliders\thanks{Summary report of the 
Top Quark Working Group
at the Workshop on Physics at the First Muon Collider and at the
Front End of a Muon Collider, November 6-9, 1997,
Fermi National Accelerator Laboratory.}}

\author{M. S. Berger$^*$ and B. L. Winer$^{\dagger}$}
\address{$^*$Department of Physics, Indiana University,
Bloomington, Indiana 47405\\
$^{\dagger}$Department of Physics, Ohio State University,
Columbus, Ohio 43210}

\maketitle

\begin{abstract}
The top quark will be extensively studied at future muon colliders. The 
threshold cross section can be measured precisely, and the small beam energy
spread is especially effective at making the measurement useful. We report
on all the activities of the top quark working group, including talks on 
top quark physics at other future colliders.
\end{abstract}

\section*{Introduction}
The top quark is expected to be more sensitive than the lighter quarks
to new physics effects.
It is also the least accessible of the quarks due to its large mass.
New colliders are under consideration that could considerably improve our
understanding of the top quark.

The fact that the top quark is heavy and there are no heavier quarks (at 
least probably not) lends some credence to the idea that the top quark 
is special.  Perhaps it is involved in the dynamics of electroweak 
symmetry breaking, or is subject to some new dynamics. Its Yukawa coupling
is comparable in size to the gauge couplings and hence has a significant
impact on the evolution of parameters with scale, and is a crucial 
ingredient in comparisons of weak-scale parameters with possible 
grand-unified theories. So it is important to measure the mass, couplings, 
and partial widths of the top quark as well as search for resonances in the 
$t\overline{t}$ spectrum. Any deviation from SM expectations would be 
of great interest.
 
Among the issues of paramount importance are (i) the nature of the
absence of flavor changing neutral currents. This can be understood
in the Standard Model (SM) with one Higgs boson as arising from the 
simplicity of the Model. Only one Higgs doublet does not allow 
tree-level effects (via the GIM mechanism)
which are naturally there in almost any extension
of the standard model. 
(ii) The test of QCD in a new regime, and the accurate measurement of 
its mass and couplings.

We summarize here the activities of the Top Quark Working Group\cite{group}.
Discussions highlighted the potential of muon colliders, while there were 
additional discussions on future electron-positron and hadron colliders. 
We refer the interested reader to the many recent 
reviews\cite{review1,review2,review3,review4,review5}
of top quark physics for a more comprehensive treatment.

\section*{Top-quark Mass Measurement at the
$\mu^+\mu^-\to \lowercase{t \bar t}$ Threshold}
One attractive feature of lepton colliders is the ability to do threshold 
cross section 
measurements. The $W$ boson mass has been determined at LEP II by measuring
the cross section $e^+e^-\to W^+W^-$ at the center-of-mass energy 
$\sqrt{s}=161$~GeV. In general, accurate measurements of particles masses, 
couplings and widths are possible by measuring production cross sections 
near threshold. 
The possibility of measuring the top quark mass as well as other relevant
parameters at a Next Linear Collider (NLC) has been under discussion for 
some years and was nicely review for the group by Raja\cite{raja}. 
This technique has also been investigated more recently
in the context of muon colliders.
There is very rich physics associated with the $t\bar t$ threshold, including
the determination of $m_t$, $\Gamma_t$ ($|V_{tb}|$), $\alpha_s$, and
possibly $m_h$ \cite{kuhn}.
Ref.~\cite{raja} contains a nice review of the most salient experimental 
measurements that can be done. This includes not only measurements of the 
mass and couplings of the top quark from the total cross section, but also
extracting information from the various distributions of the 
detected particles. These issues carry over completely to the
muon collider case, with some important differences in the characteristics 
of the collider beam having some impact on the sensitivities (see below).
 
Fadin and Khoze first
demonstrated that the top-quark threshold cross section is
calculable since the large
top-quark mass puts one in the perturbative regime of QCD,
and the large top-quark width effectively screens nonperturbative effects
in the final state \cite{fk}.
Such studies have since been performed by several
groups \cite{feigenbaum,kwong,sp,jht,bagliesi,sfhmn,immo,fms}.
The phenomenological potential 
is given at small distance $r$ by two-loop perturbative QCD and
for large $r$ by a fit to quarkonia spectra.

\begin{figure}[htb]
\leavevmode
\begin{center}
\vspace*{-0.1in}
\epsfxsize=3.0in\hspace{0in}\epsffile{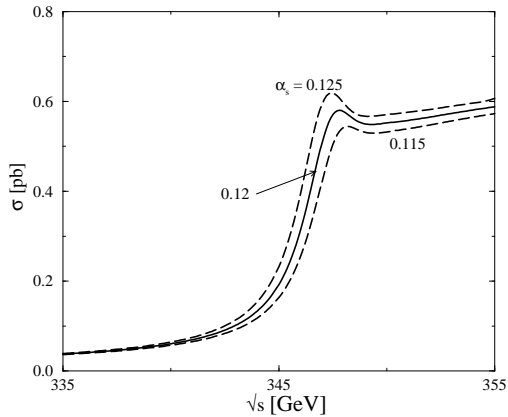}
\vspace*{-0.1in}
\end{center}
\caption[]{\footnotesize\sf The cross section for $\mu^+\mu^- \to t \bar{t}$
production in the threshold region, for $m_t=175$ GeV and
$\alpha_s(M_Z)=0.12$ (solid) and 0.115, 0.125 (dashes).
Effects of ISR and beam smearing are included. This
figure is from Ref.~\cite{bbgh3}.}
\label{ttfigure}
\end{figure}

The most important parameters affecting the shape of the threshold cross 
section are the top quark mass $m_t$ and the strong coupling constant 
$\alpha_s$. The mass determines at what energy the threshold turns on, while
the strong coupling determines the binding between the $t\overline{t}$ pairs
and hence causes in principle a resonance structure in the spectrum.
However, since the top quark mass has turned out to be so large, only the
$1S$ state appears as a structure on the threshold curve. The stronger
the strong coupling, the tighter the binding and the lower the $1S$ peak
occurs in energy. Weaker coupling also smooths out the threshold peak.
These effects are illustrated in Fig.~\ref{ttfigure}.
Clearly the effects of varying $m_t$ and $\alpha _s$ are
correlated; this fact necessitates that some kind of scan be performed 
of the threshold cross section.

The scan can be optimized in various ways depending on the parameters
one is most interested in measuring. In addition there is information 
contained in the various distributions of the final state particles.
The momentum distribution of the top quark pairs as well as the 
forward-backward asymmetry are sensitive to the top 
quark width and $\alpha _s$.

\begin{figure}[htb]
\leavevmode
\begin{center}
\vspace*{-0.1in}
\epsfxsize=3.0in\hspace{0in}\epsffile{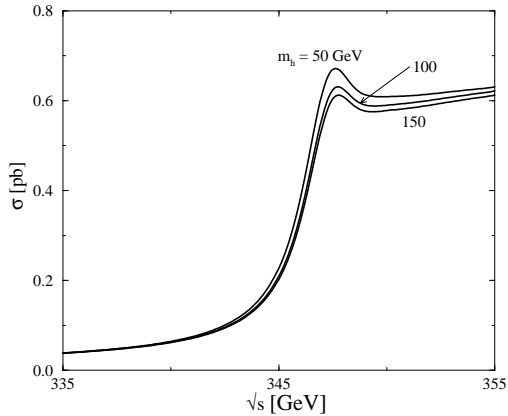}
\vspace*{-0.1in}
\end{center}
\caption[]{\footnotesize\sf The dependence of the threshold region
on the Higgs mass, for $m_h=50, 100, 150$~GeV. Effects of
ISR and beam smearing have been included,
and we have assumed $m_t=175$~GeV and $\alpha_s(M_Z)=0.12$. This
figure is from Ref.~\cite{bbgh3}.}
\label{figure8}
\end{figure}

The presence of the Higgs boson affects the threshold curve. This contribution
depends on the Higgs boson mass and the Yukawa coupling with which the
Higgs boson couples to the top quark. The Yukawa coupling is fixed in 
the Standard Model for a given top quark mass, but could be different in
extensions to the Standard Model. The Higgs boson contribution mainly affects
the overall normalization of the threshold curve. 
Since the exchange of a light Higgs boson can affect the threshold shape,
a scan of the
threshold cross section can in principle yield some information about the Higgs
mass and its Yukawa coupling to the top quark.
Figure~\ref{figure8} 
shows the dependence of the threshold curve on the Higgs mass, $m_h$.
However, it may be difficult to disentangle such a Higgs
effect from two-loop QCD effects, which are not yet fully
calculated~\cite{hoang}. 
Since one does not expect an accurate measurement of the Higgs mass 
from the $t\overline{t}$ threshold, one should properly think of the 
Higgs contribution as a systematic uncertainty that can be removed
by measuring the Higgs mass and top quark Yukawa coupling elsewhere.

Beamstrahlung is the emission of radiation by one beam due to the action of 
the effective magnetic field of the other beam. This is expected to be 
an important issue at electron-positron colliders and clearly depends 
on the  machine design. Muon colliders are expected to naturally have 
negligible beamstrahlung due to the large mass of the muon.

A more minor difference between the electron colliders and the muon colliders
is the difference in the amount of initial state radiation (ISR).
The expansion parameter is 
\begin{eqnarray}
&&\beta ={{2\alpha }\over \pi}\left (\ln (s/m_\ell^2)-1\right )\;,
\end{eqnarray}
where $m_\ell$ is the mass of the initial state particle (the electron 
or the muon). The radiator function that must be convoluted with the 
underlying cross section is
\begin{eqnarray}
{\cal D}(x)&=&1+{2\alpha\over \pi}(\pi^2/6-1/4)
\left[\beta x^{\beta-1} \left(1+ {3\over 4}\beta\right)
-\beta \left(1-{x\over 2}\right)\right]\nonumber
\end{eqnarray}
The ISR is reduced somewhat at a muon collider relative to
an electron collider.

Two methods have been used to calculate the threshold cross section. The
first (the coordinate-space approach) 
involves solving a nonrelativistic Schr\"{o}dinger equation that splices
together a QCD potential from perturbative QCD at small distance scales
with one that is derived from fits to quarkonia spectra. The other method 
(the momentum space approach) involves solving a Bethe-Salpeter equation.
The construction and the relationship between the QCD potentials used in 
each case is subject of recent study\cite{jkpst}.
At a high luminosity muon collider it is evident that theoretical uncertainties
in the threshold cross section might be the limiting factor in 
the ultimate obtainable precision.

The calculations used in simulations so far have been done mostly at the
next-to-leading order (NLO) level. These next-to-next-to-leading order (NNLO)
have not been taken into account even though there contributions can be
important. For example, the ${\cal O}(\alpha_s^2)$ relativistic corrections 
can shift the location of the $1S$ peak by $m_t\alpha_s^4\sim 150$~MeV 
and introduce a shift in the normalization of the total cross section 
of order $\alpha_s^2\sim 3\%$\cite{hoang1}. Hoang described a procedure
to calculate some of the NNLO corrections to the threshold cross section
using NRQCD, an effective field theory of QCD for heavy quarks. NRQCD
does away with the need for a phenomenological potential, and allows at
least in principle the calculation of the cross section and all
the distributions from the QCD Lagrangian. 

\begin{figure}[htb]
\leavevmode
\begin{center}
\vspace*{-1.2in}
\epsfxsize=3.0in\hspace{0in}\epsffile{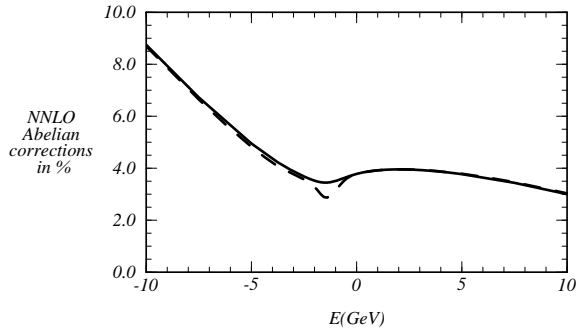}
\vspace*{-1.2in}
\end{center}
\caption[]{\footnotesize\sf The NNLO Abelian corrections to the cross section
for $\Gamma _t=1.56$~GeV (solid line) and 0.80~GeV (dashed line), from 
Ref.~\cite{hoang1}.}
\label{hoangfig}
\end{figure}

The Abelian part (i.e. those contributions also present in QED) of the
NNLO corrections calculated by Hoang are shown in Fig.~\ref{hoangfig}. Notice 
that the corrections are a few percent and is fairly constant for the part
of the cross section including and above the $1S$ peak, $E>-5$~GeV. (Compare
the location of the peak in Fig.~\ref{ttfigure}.)

Muon colliders are expected to naturally have a small spread in 
beam energy making them an ideal place 
to study the excitation curve. We present now the parameter determinations 
that are possible from measuring the total cross section near threshold at a 
$\mu ^+\mu ^-$ collider\cite{bbgh3,ttbar,berger}.

The beam energy spread 
at a $\mu^+\mu^-$ collider is expected to naturally be small. 
The rms deviation $\sigma $ in $\sqrt{s}$ is given by \cite{bbgh1,bbgh2}
\begin{equation}
\sigma = (250~{\rm MeV})\left({R\over 0.1\%}\right)\left({\sqrt s\over {\rm
350\ GeV}}\right) \;,
\end{equation}
where $R$ is the rms deviation of the Gaussian beam profile.
With $R\alt 0.1\%$ the resolution $\sigma$ is of the same
order as the measurement one hopes to make in the top mass.
For $t\overline{t}$ studies the exact shape of the beam is not
important if $R\alt 0.1\%$. We take
$R=0.1\%$ here; the results are not improved significantly with better
resolution\footnote{The most recent TESLA design
envisions a beam energy spread of $R=0.2\%$\cite{miller}, and a high 
energy $e^+e^-$ collider in the large VLHC tunnel would have a beam 
spread of $\sigma _E=0.26$~GeV\cite{norem}.}.

\begin{figure}[htb]
\leavevmode
\begin{center}
\vspace*{-0.1in}
\epsfxsize=3.0in\hspace{0in}\epsffile{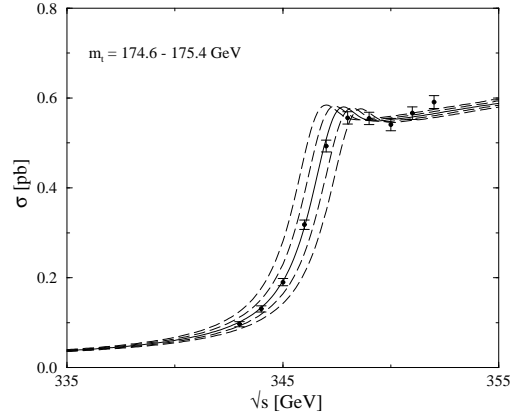}
\vspace*{-0.1in}
\end{center}
\caption[]{\footnotesize\sf Sample data for $\mu^+ \mu^-
\rightarrow t \bar{t} $ obtained assuming a scan over
the threshold region devoting 10~fb$^{-1}$ luminosity to each data point.
A detection efficiency of 29\% has been assumed \cite{fms}
in obtaining
the error bars. The threshold curves correspond to shifts in $m_t$ of 200~MeV
increments. Effects of ISR and beam smearing have been included,
and the strong coupling $\alpha_s(M_Z)$ is taken to be 0.12. This
figure is from Ref.~\cite{bbgh3}.}
\label{figure6}
\end{figure}

\begin{figure}[htb]
\leavevmode
\begin{center}
\vspace*{-0.1in}
\epsfxsize=3.0in\hspace{0in}\epsffile{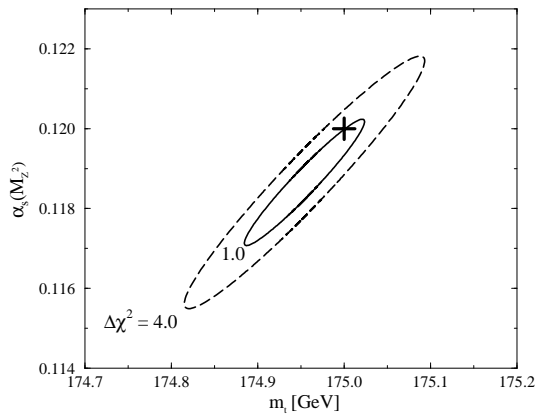}
\vspace*{-0.1in}
\end{center}
\caption[]{\footnotesize\sf The $\Delta \chi ^2=1.0$ and
$\Delta \chi ^2=4.0$ confidence
limits for the sample data shown in Fig.~\ref{figure6}. The ``+''
marks the input values ($m_t=175$~GeV and $\alpha_s(M_Z)=0.12$) 
from which the data were generated. This
figure is from Ref.~\cite{bbgh3}.}
\label{figure7}
\end{figure}

Suppose one starts with the nominal values of $m_t=175$ GeV
and $\alpha_s(M_Z)=0.12$. Assuming that 10~fb$^{-1}$ integrated luminosity
is used to measure the cross section at each energy in
1~GeV intervals, one can imagine obtaining the hypothetical sample data, 
shown in Fig.~\ref{figure6}. Cuts must be performed to eliminate the 
backgrounds; following Ref.~\cite{fms} a 29\% detection efficiency has 
been assumed for the signal where the $W$'s decay hadronically\footnote{This
efficiency includes the decay branching fraction.}
The data points can then be fit to theoretical predictions
for different values of $m_t$ and $\alpha_s(M_Z)$; the
likelihood fit that is obtained is shown
as the $\Delta \chi ^2$ contour plot in Fig.~\ref{figure7}.
The inner and outer curves are the $\Delta \chi^2=1.0$ (68.3\%) and $4.0$
(95.4\%)
confidence levels respectively for the full 100~fb$^{-1}$ integrated
luminosity.
Projecting the $\Delta \chi^2=1.0$ ellipse on the $m_t$ axis,
the top-quark mass
can be determined to within $\Delta\mt\sim 70\mev$,
provided systematics are under control. 
With an integrated luminosity of 10~${\rm fb}^{-1}$,
the top-quark mass can be measured to 200~MeV.

QCD measurements at future colliders and
lattice calculations will presumably determine
$\alpha_s(M_Z)$ to 1\% accuracy (e.g.\ $\pm 0.001$) \cite{alphas}
by the time muon colliders are constructed so the uncertainty in
$\alpha_s$ will likely be similar to
the precision obtainable at a $\mu^+\mu^-$ and/or $e^+e^-$
collider with 100~fb$^{-1}$ integrated luminosity.
If the luminosity available for the threshold measurement is
significantly less than 100~fb$^{-1}$,
one can regard the value of $\alpha_s(M_Z)$ coming from
other sources as an input, and thereby improve the
top-quark mass determination.

There is some theoretical ambiguity in the
mass definition of the top quark. The theoretical
uncertainty on the quark pole mass
due to QCD confinement effects is of  order
$\Lambda_{QCD}$, {\it i.e.},  a few hundred MeV \cite{sw,smith}.
For example, this theoretical ambiguity manifests itself in relating 
quark pole mass to other definitions of
the top quark mass (such as the running top quark mass, $\overline{m}_t(\mu)$) 
that might be relevant as input to radiative correction
calculations. So it is not
clear that an extraction of the top-quark mass better than $\Lambda_{QCD}$
is useful, at 
least at the present time.

Systematic errors in experimental efficiencies are not a significant
problem for the $t\anti t$ threshold determination of $\mt$.  This can
be seen from Fig.~\ref{figure6}, 
which shows that a 200 MeV shift in $\mt$ corresponds
to nearly a 10\% shift in the cross section
on the steeply rising part of the
threshold scan, whereas it results in almost no change in $\sigma$
once $\rts$ is above the peak by a few GeV.  Not only will efficiencies
be known to much better than 10\%, but also systematic uncertainties
will cancel to a high level of accuracy in the ratio of
the cross section measured above the peak to measurements
on the steeply-rising part of the threshold curve.

Differences of cross sections at energies
below, at, and above the resonance peak, along with the location of the
resonance peak, have different dependencies on the parameters
$m_t$, $\alpha_s$, $m_h$ and $|V_{tb}|^2$
and should allow their determination. Consequently, the scan 
procedure described here can be further optimized for extraction
of a particular parameter\cite{bbgh3}.

\section*{Top quark pairs above threshold}

The production of top quark pairs at energies above the threshold region 
at muon colliders will provide a great opportunity for searching for anomalous 
couplings and rare decays of the top quark. The production of top quarks
can be used to test couplings to the neutral gauge bosons $\gamma, Z$ in the
production and to the $W$ in the decays. An important new feature of top
quark decays is the fact that the top quark decays before it has a chance
to hadronize, so the spin information can be preserved from production to 
decay. This introduces the possibility that spin correlations between 
$t$ and $\overline{t}$ might even be measurable\cite{ps}. 
Finally a large sample of top quark pairs allows us to search for possible 
rare decays, e.g. decays into a charged Higgs boson or flavor changing
decays ($t\to c$) which are exceedingly small in the Standard Model.

\begin{figure}[htb]
\leavevmode
\begin{center}
\vspace*{-0.5in}
\epsfxsize=3.0in\hspace{0in}\epsffile{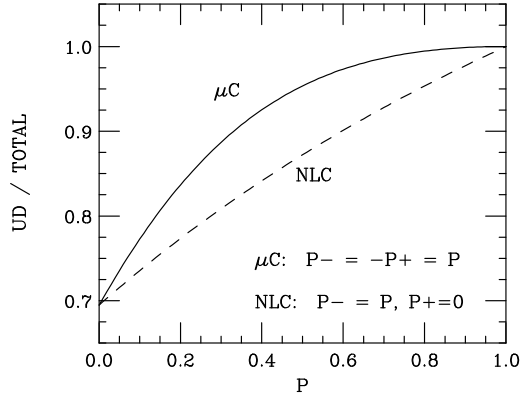}
\vspace*{-1.5in}
\end{center}
\caption[]{\footnotesize\sf Fraction of the total cross section in the off-diagonal basis' 
Up-Down spin configuration as a function of the polarization. 
Both beams are assumed to be polarized for the Muon Collider ($\mu$C)
but only one beam for the NLC.
The plot is taken from Ref.~\cite{parke}.}
\label{parkefig}
\end{figure}

The off-diagonal basis described by Parke\cite{parke}
is superior to the standard helicity basis and allows 
one to describe the $t\overline{t}$ in their simplest possible
terms. The basis is characterized by a spin angle $\xi$ between the 
$t$ spin and the $\overline{t}$ momentum given by
\begin{eqnarray}
&&\tan \xi={{(f_{LL}+f_{LR})\sqrt{1-\beta^2}\sin \theta ^*}\over
{f_{LL}(\cos \theta ^*+\beta)+f_{LR}(\cos \theta ^*-\beta)}}\;,
\end{eqnarray}
where $\beta$ is the top quark velocity, $\theta ^*$ is the top quark 
scattering angle, and $f_{IJ}$ is a combination of muon (or electron) and 
top quark couplings (see Ref.~\cite{ps}). This spin basis interpolates 
between the beam direction at threshold and the top quark direction very 
far above threshold.
Polarization of the incoming beams enhances the sensitivity
to the basis choice. The dominant spin component's 
fraction of the total as function of the polarization of the beams is plotted
in Fig.~\ref{parkefig}
Two different machines are included: the
muon collider is assumed to have equal
but opposite polarization for the $\mu^+$ and $\mu^-$ beams; the NLC has
only the electron beam polarized. 
A muon collider can do as well as an electron-positron machine with 
relatively less polarization.

In the future one hopes that detailed studies
of the effects of anomalous coupling in the off-diagonal basis
will become available. QCD corrections have  
been calculated and shown to be small\cite{KNP}.
If the muon or electron beams can be polarized, 
the sensitivity then to anomalous
couplings can be enhanced\cite{parke}.

\begin{figure}[htb]
\leavevmode
\begin{center}
\vspace*{-1.2in}
\epsfxsize=3.0in\hspace{0in}\epsffile{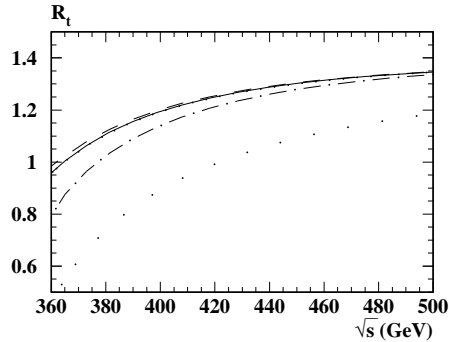}
\vspace*{-1.2in}
\end{center}
\caption[]{\footnotesize\sf The total normalized photon-mediated 
cross section at the two-loop
level versus $\sqrt{s}$ for the renormalization scales $\mu=M_t$
(dashed), $\mu=2 M_t$ (solid) and $\mu=\sqrt{s}$ (dotted line),
$M_t=175$~GeV and $\alpha_s^{(5)}(M_z)=0.118$. For
comparison also the Born (wide dots) and the one-loop cross section
for $\mu=2 M_t$ (dashed-dotted line) are displayed. To improve the
stability under renormalization scale variations the known three-loop
${\cal{O}}(\alpha_s^3)$ in the large momentum
expansion have been added to the two-loop cross section.
The plot is taken from Ref.~\cite{hoang2,chetyrkin}.}
\label{hoangfig2}
\end{figure}

Hoang described progress in two-loop calculations of the top production
cross section in the kinematic region above the threshold\cite{hoang2}.
His results for the part of the cross section do not yet include the 
axial pieces, but the one can see the improvement in the stability under
variations of the renormalization scale $\mu$ in Fig.~\ref{hoangfig2}.

\section*{Flavor changing neutral currents}

There are strong phenomenological constraints on flavor changing neutral 
currents (FCNC) from $K$ physics, for example. However, the possibility of
FCNC in top couplings remains to be explored. Flavor changing decays of the 
top quark (e.g. $t\to c\gamma $) are extremely suppressed in the 
Standard Model, but new physics contributions could enhance the rate 
(see below and Ref.~\cite{tao}). 
Another possible source of FCNC is to look for decays of 
Higgs bosons into single top quarks.

A unique feature of the muon collider is the ability to produce the 
Higgs boson in the $s$-channel\cite{bbgh1,bbgh2}. This possibility 
arises because the Higgs boson coupling to leptons is proportional
to their mass. Since  the Higgs boson might be a very narrow object
if one can center muon beams with very sharp beam profiles on
the resonance energy one could produce a substantial sample 
of Higgs bosons and study their decays. An interesting set of 
decays are flavor changing ones, which for the Standard Model 
Higgs boson are completely absent. Many extensions to the 
Standard Model have flavor changing processes at the tree-level, 
and then the question
becomes why they are so small in the physics that we see.
If the Higgs boson is heavier than the top quark then one can
consider the possibility that the Higgs boson decays via
$H\to t\overline{c} + c\overline{t}$. 

Reina and collaborators\cite{ars,arsmumu} 
chose a particular two-Higgs doublet model to 
provide examples of the kind of effects one might potentially see at a muon
collider. The model is given by the Lagrangian
\begin{eqnarray}
&&{\cal L}_Y^{(III)}=\eta_{ij}^U\overline{Q}_{i,L}\tilde{\phi}_1U_{j,R}
+\eta_{ij}^D\overline{Q}_{i,L}\phi_1D_{j,R}
+\xi_{ij}^U\overline{Q}_{i,L}\tilde{\phi}_2U_{j,R}
+\xi_{ij}^D\overline{Q}_{i,L}\phi_2D_{j,R}+h.c.\;,
\end{eqnarray}
where $\eta $ and $\xi$ are non-diagonal Yukawa matrices. Usually at this 
point one imposes a discrete symmetry to eliminate tree-level FCNCs.
Instead a reasonable choice is to assume that the flavor changing couplings 
adhere to the same hierarchy as the fermion masses\cite{hier}
\begin{eqnarray}
&&\xi_{ij}= \lambda_{ij}{{\sqrt{m_im_J}}\over v}\;,
\end{eqnarray}
and tree-level FC couplings can be substantial for the top quark.
This model (like all two Higgs doublet models) is parameterized by 
a mixing angle $\alpha$ between the two neutral scalars. 

An important consideration for producing Higgs bosons in the $s$-channel,
is the relative size of the beam width to the width of the Higgs 
boson\cite{bbgh1,bbgh2}.
A sufficiently sharp beam, if suitably tuned to the resonance energy, can 
take full advantage of the resonant cross section.
One can define the effective
cross section as the convolution of the Breit-Wigner
$\sigma^{BW}_{tc}$ cross section with a gaussian beam energy spread,

\begin{equation}
\sigma^{eff}_{tc}=\int d\sqrt{s^\prime}
\frac{\mbox{exp}[-(\sqrt{s^\prime}-\sqrt{s})^2/2\sigma^2]}
{\sqrt{2\pi}\sigma}\,\sigma^{BW}_{tc}(s^\prime)\,\,\,\,,
\label{E:lreina:14}
\end{equation}
where the rsm of the gaussian distribution is defined in
terms of the parameter $R$.

In the analysis in Ref.~\cite{reina} the effective cross section after 
convoluting with the beam width is expressed in units of $R_{\mu\mu}$ 
(not to be confused with the parameter $R$ describing the beam width) as 
follows
\begin{equation}
R_{tc}=\frac{\sigma^{eff}_{tc}}{\sigma_0}=
R({\cal H})\left(B({\cal H}\rightarrow \bar t c) + 
B({\cal H}\rightarrow\bar c t)\right)\,\,\,\,,
\label{E:lreina:18}
\end{equation}
\noindent where
$\sigma_0\!=\!\sigma(\mu^+\mu^-\rightarrow\gamma\rightarrow e^+e^-)$
and $R({\cal H})\!=\!\sigma_{\cal H}/\sigma_0$ for $\sigma_{\cal H}$
the total cross section for producing ${\cal H}$. 

\begin{figure}[htb]
\leavevmode
\begin{center}
\epsfxsize=3.0in\hspace{0in}\epsffile{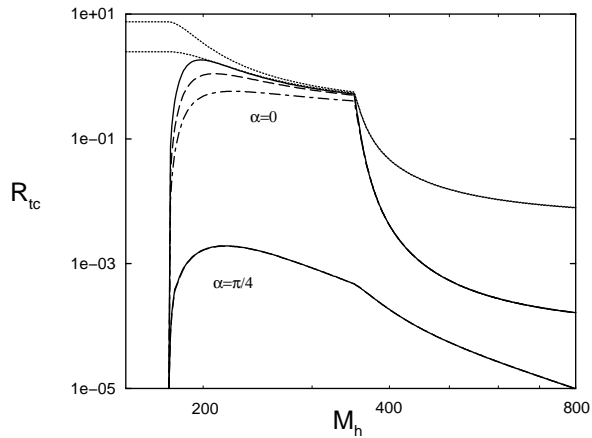}
\end{center}
\caption[]{\footnotesize\sf The value of $R(h^0)$ is 
shown as a function of $M_{h^0}$
in a pure Breit-Wigner case (upper dotted line) and when the gaussian
width distribution of the beam energy (for $R\!=\!0.01$) is taken into
account (lower dotted line).  The ratio $R_{tc}$ is also shown for
different values of the resolution parameter $R\!=\!0.001$ (solid),
0.01 (dashed) and 0.03 (dot-dashed), when $\alpha\!=\!0$ (upper group
of curves) and when $\alpha\!=\!\pi/4$ (lower group of curves). This
figure is from Ref.~\cite{reina}.}
\label{reina}
\end{figure}

As described above the results depend on how $\Gamma_{h^0}$ 
compares to the resolution parameter $R$. 
This is shown in Fig.~\ref{reina} which shows the results for 
the pure Breit-Wigner as well as different assumptions for $R$.
Two choices for the mixing angle $\alpha $ shows the kind of variations one
can get ($\alpha =0$ means the $s$-channel Higgs $h^0$ does not couple to 
the gauge bosons which give competing decay channels $h^0\to W^+W^-,ZZ$.

With 1~fb$^{-1}$ of integrated luminosity and gets around 100 
$t\overline{c}+c\overline{t}$ events for $\alpha =0$ and a few for 
$\alpha =\pi/4$\cite{reina}. Clearly higher luminosity would be 
advantageous here.

Flavor changing rare decays of the top occur at an exceeding small rate
in the Standard Model and the Minimal Supersymmetric Model (MSSM).
However if R-parity violation occurs, then there is at least some 
hope that rare decays $t\to c$ could be detected at the upgraded 
Tevatron\cite{yang}.
 
R-parity violation also could give an additional source of single 
top producation. For example the lepton-number violating 
coupling\footnote{For a definition and discussion of the
R-parity violating couplings, see Refs.~\cite{datta,oakes}.} 
$\lambda ^\prime$ gives
rise to the $s$-channel process
\begin{eqnarray}
u\overline{d}\to \tilde{\ell}\to t\overline{b}\;.
\end{eqnarray}
The baryon-number violating coupling $\lambda ^\prime\prime$ gives rise
to the $s$-channel processes
\begin{eqnarray}
c\overline{d}\to \tilde{s}\to t\overline{b}\;, \\
c\overline{s}\to \tilde{d}\to t\overline{b}\;,
\end{eqnarray}
and the $t$-channel process
\begin{eqnarray}
u\overline{d}\to t\overline{b}\;.
\end{eqnarray}
The prospects for setting bounds on R-parity violation from these processes
is discussed in Refs.~\cite{yang,datta,oakes}.

\section*{Gluon radiation}

Top quark production involves gluon radiation because the top quark is 
a colored particle. In a hadron collider the gluon can arise in the 
initial and final states\cite{orr_etal}. 
As far as gluon radiation at lepton colliders 
is concerned, there is no significant difference
between $\mu^+\mu^-$ and $e^+e^-$ colliders. There is no gluonic
ISR, but the radiation must still be divided into production and decay
stage radiation, i.e. the gluon can be thought of as originating off the
produced $t$ or $\overline{t}$, or it can be thought of as being 
among the decay products of the $t$ or $\overline{t}$. 
Gluon radiation needs to be 
understood if we are able to do precise momentum reconstructions to obtain 
$m_t$, and also to identify top events by using mass cuts\cite{orr}.

\begin{figure}[htb]
\leavevmode
\begin{center}
\vspace*{-1.0in}
\epsfxsize=3.0in\hspace{0in}\epsffile{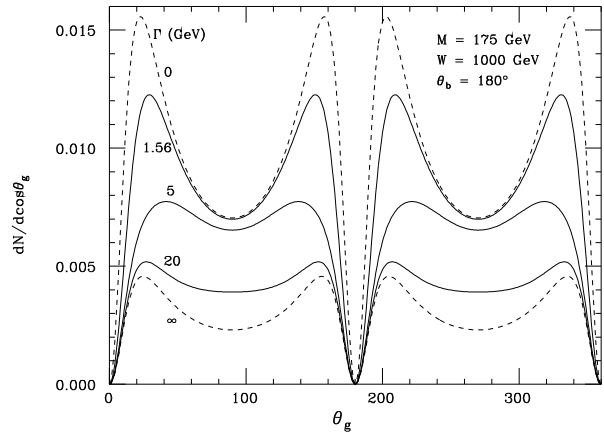}
\vspace*{-1.0in}
\end{center}
\caption[]{\footnotesize\sf Soft gluon distribution in top production 
and decay at lepton colliders 
as described in the text, for $t$'s decaying to backward $b$'s and 
collision energy $1\ {\rm TeV}$, from Ref.~\cite{orr}.}
\label{orrfig}
\end{figure}

What one would really like to do is study the gluon radiation pattern.
There are interference effects between the production and decay stage 
radiation that is potentially sensitive to the top quark width $\Gamma _t$. 
This occurs when the gluon energy is comparable to $\Gamma _t$.
One such
radiation pattern is shown in Fig.~\ref{orrfig} for a particular kinematic
configuration and a variety of values for $\Gamma _t$.

Whether one can really extract information about the top quark width 
by observing the interference patterns remains an open question\cite{orr}.
The distributions that would be seen at a muon collider would involve
all possible kinematic configurations.
Moreover, the interference effects are much smaller for a 500~GeV collider
which is more likely to be available first.
Nevertheless, the interference is an interesting feature
of QCD and one would like to see it even if it does not offer a new
way to measure $\Gamma _t$.

\section*{Top quark physics at future hadron colliders}

Hughes\cite{hughes} summarized top quark physics at future hadron colliders
for the working group. He compared and contrasted the expected measurements
at Run I of the Tevatron with what is expected at Run II, Tev33, and the 
LHC.

A possible manifestation of new physics is resonance structure in the 
$t\overline{t}$ mass distribution. The Tevatron is sensitive to both color
singlet and color octet, while at the LHC should be relatively
insensitive to new color singlet gauge bosons (like $Z^\prime $) because
gluon fusion is the dominant means of producing $t\overline{t}$ pairs.

Rare top decays ($t\to c$) are expected to be very small in the Standard
Model due to the GIM mechanism and the competing (Cabibbo-allowed) decay
$t\to b$. Branching fractions are typically of order $10^{-10}$ in the 
Standard Model, so the observation of rare
decays of this sort indicates new physics. With 2~fb$^{-1}$, one should be
able to get to the $10^{-3}$ level for $t\to c\gamma$\cite{hughes}.

Observing single top production will become possible after the Tevatron 
upgrade. The single top production is proportional to the partial width
$\Gamma(t\to Wb)$ and provides an expected precision of 16\% for Run II and
9\% for Tev33\cite{hughes}. 
Since the final state is a $Wjj$ configuration rather than 
a $Wjjjj$ one, the background from QCD processes should be much higher
than the usual $t\overline{t}$ case\cite{tao}. 
With 2~fb$^{-1}$, one should have 
roughly 100 events in the final sample.  

\section*{Detector and Background}
Roser\cite{roser} described progress on a strawman detector for a 
muon collider and the interplay
between the machine designers and detector designers. The detector backgrounds
are actually more under control for the high energy (4~TeV) collider where
most of the particles continue down the beam pipe without ever entering the
dectector.
 
While the total ``background energy'' does not depend on the energy of the 
beam, the ``visible'' background will depend on details of the lattice and the
detector itself. There will be the usual collider backgrounds, such as beam 
halo, beam-beam, etc. Since backgrounds at a muon collider are expected to 
be large, one will want to use a large number of detector channels to achieve 
reasonable occupancies.

The following techniques are being used to study the detector: (1) lattice
simulators COSY and MAD, (2) detector simulators MARS and GEANT, and 
(3) event generators LUND, PYTHIA, etc.

GEANT simulations yield radial particle fluxes per crossing for a layer
of silicon at a radius of 10~cm:
\begin{eqnarray}
750\:  {\rm photons/cm}^2 \to &&2.3\:  {\rm Hits/cm}^2\nonumber \\
110\:  {\rm neutrons/cm}^2 \to &&0.1\:  {\rm Hits/cm}^2\nonumber \\
1.3\:  {\rm charged\: tracks/cm}^2 \to &&1.2\:  {\rm Hits/cm}^2\nonumber \\
{\rm Total} \to &&3.7\:  {\rm Hits/cm}^2\nonumber 
\end{eqnarray}
This translates into a 0.4\% occupance in $300 \times 300~\mu$m$^2$ pixels.
The corresponding numbers at a radius of 5~cm are 13.2~Hits/cm$^2$ for a 1.3\%
occupancy. The radiation dose in the silicon vertex detector at a 4~TeV muon
collider at a radius of
10~cm is comparable to that of the LHC operating at 
$10^{34} {\rm cm}^{-2}{\rm s}^{-1}$ luminosity.

Backgrounds come from synchrotron radiation from electrons and from 
electromagnetic showering close to the detector. The lattice focuses the muons
at the interaction point, so that the electrons cannot be kept in the beam
pipe. 

In summary, technology choices and detector optimization will require 
dedicated work. Because of the large backgrounds arising from 
the decaying muons, there will be pressure to compromise on $4\pi$ coverage.

\section*{Conclusion}

An important issue regarding top physics at a muon collider
is the amount of luminosity that would be available. Most of the 
signals presented here require significantly more than 1~fb$^{-1}$ of 
integrated luminosity. The years ahead promise to be very exciting 
as we become able to study the top quark properties in more detail.

\section*{Acknowledgments}

We thank all the working group participants\cite{group} 
for their contributions and discussion, and
the conference organizers for an efficient and interesting workshop.

\end{document}